\documentclass[%
 reprint,
 amsmath,amssymb,
 aps,
prl,
]{revtex4-2}

\usepackage{graphicx}
\usepackage{dcolumn}
\usepackage{bm}
\usepackage{xcolor} 
\usepackage{algorithm}
\newcommand{\iu}{\mathrm{i}}

\begin{document}

\preprint{APS/123-QED}

\title{A multi-ensemble mean-field reduction method for networks of globally coupled phase oscillators with arbitrary parameter distributions}

\author{Richard Gast$^{1}$, Shotaro Takasu$^{1}$, Helmut Schmidt$^{2+}$ Ann Kennedy$^{1+\dagger}$}
\affiliation{$^1$ Department of Neuroscience, Scripps Research, San Diego, US.}
\affiliation{$^2$ Institute of Computer Science, Czech Academy of Sciences, Prague, CZ.}
\thanks{$^+$These authors contributed equally to this work}
\thanks{$^\dagger$Corresponding author: Ann Kennedy (\texttt{akennedy@scripps.edu})}


\date{\today}

\begin{abstract}
Understanding the dynamical properties of coupled phase oscillator systems with heterogeneous oscillator frequencies has been a long-standing challenge of complex systems theory.
While the seminal work of Ott and Antonsen dramatically improved our theoretical understanding of coupled phase oscillators for a small family of oscillator frequency distributions, we here present a mean-field reduction method for arbitrary frequency distributions. 
Our method leverages the drastic dimensionality reduction obtained for Lorentzian frequency distributions, and combines it with a data-driven multi-ensemble approach.
As such, the method renders the Ott-Antonsen equations directly applicable to empirical distributions of phase oscillator frequencies, often achieving a drastic dimensionality reduction and allowing to study real-world physical and biological systems by means of stability, sensitivity, and bifurcation analyses.
\end{abstract}

\maketitle


\textit{Introduction ---} Coupled phase oscillators have been broadly applied as a mathematical framework for studying physical, biological, and chemical systems with interacting periodic processes, such as power grids \cite{olmi_hysteretictransitionskuramoto_2014,molnar_asymmetry_2021,sajadiSynchronizationElectricPower2022}, chemical reactions \cite{kuramoto_patternformationoscillatory_1976,epstein_oscillatingchemicalreactions_1983,kuramotoChemicalTurbulence1984,ngonghala_extrememultistabilitychemical_2011}, or neural networks \cite{luke_complete_2013,wedgwood_phaseamplitudedescriptionsneural_2013,stiefel_neuronsoscillators_2016,ashwin_mathematicalframeworksoscillatory_2016,saha_extremeeventsfitzhughnagumo_2017,pietrasNetworkDynamicsCoupled2019}.
As individual oscillators are generally non-identical, it is a key question across all these systems, how heterogeneity in the intrinsic oscillator frequencies affects the system dynamics. 
In this letter, we introduce a mean-field method for studying the dynamics of phase oscillator systems with arbitrary frequency distributions, which naturally lends itself to the study of oscillator heterogeneity.

We consider networks of globally coupled phase oscillators, where the phase of each oscillator evolves according to
\begin{align}
    \dot \theta_i = \omega_i + \frac{K}{N} \sum_{j=1}^N G(\theta_j,\theta_i), \label{eq:theta}
\end{align}
and is governed by an intrinsic frequency $\omega_i$ that introduces quenched disorder at the level of the oscillators.
For the family of systems where the function $G$ takes the form
\begin{equation}
    G(\theta_j, \theta_i) = c_{-1}(\theta_j) e^{-\iu\theta_i} + c_0(\theta_j) + c_{+1}(\theta_j) e^{\iu\theta_i},\label{eq:OA_coupling}
\end{equation}
the dynamics of the network converge to the Ott-Antonsen manifold \cite{ott_low_2008,pietras_ott-antonsen_2016,bick_understanding_2020}.
The results of this letter are restricted to phase oscillator networks of the form given by Eqs.~\eqref{eq:theta} and \eqref{eq:OA_coupling}.
Note that this family includes the set of systems where phase coupling only depends on the phase difference, i.e. $G(\theta_j, \theta_i) = F(\theta_j-\theta_i)$, for which the Ott-Antonsen ansatz conditions are met for coupling functions of the form $F(\phi) = \Gamma_0 + r \sin(\phi + \beta)$ \cite{ott_low_2008,bick_understanding_2020}.

Assuming the special case of a Lorentzian distribution of intrinsic oscillator frequencies
\begin{equation}
    \rho(\omega) = \frac{\Delta}{\pi([\omega-\bar\omega]^2 + \Delta^2)}, \label{eq:lorentz}
\end{equation}
the dynamics of the network on the OA manifold can be derived from Eq.~\eqref{eq:theta}, yielding
\begin{align}
    \dot Z = (\iu\bar \omega-\Delta) Z + \frac{K}{2} Z (1-|Z|^2),
\end{align}
with $Z(t) = \frac{1}{N} \sum_i \exp(\iu\theta_i)$ representing the complex-valued Kuramoto order parameter.
This special case reduces the complexity of the network dynamics considerably, causing the network to approach either a fixed point that corresponds to a partially or completely synchronized state, or a fixed point that corresponds to an asynchronous state \cite{ott_low_2008}. 

Since the seminal work of Ott and Antonsen, it has been a widely debated question whether globally coupled phase oscillator networks with frequency distributions $\rho(\omega)$ other than Eq.~\eqref{eq:lorentz} also permit for a low-dimensional mean-field description and how their macroscopic dynamics differ from the Lorentzian case \cite{martens_exact_2009,pietras_ott-antonsen_2016,pietras_equivalence_2016,skardal_low-dimensional_2018,bick_understanding_2020,campa_study_2022}. 
Scenarios that have been studied include Gaussian, uniform \cite{skardal_low-dimensional_2018}, bimodal \cite{martens_exact_2009}, and trimodal oscillator frequency distributions \cite{pietras_equivalence_2016}, for example. 
In this letter, we consider a much broader family of oscillator frequency distributions $\rho(\omega)$, including multi-modal and non-symmetric distributions.

We provide a general mean-field reduction method for this class of networks that permits a drastic reduction in system dimensionality and enables the study of spatiotemporal pattern formation via methods from dynamical systems theory.
We demonstrate our mean-field reduction method on representative model systems and on an open-source dataset of \emph{in vitro} electrophysiological recordings from mouse neurons. 
We show that our method can (a) faithfully capture electrophysiological parameter distributions of different neuron types in different cortical layers, (b) predict the macroscopic dynamics of networks of recurrently coupled spiking neurons governed by those parameter distributions, and (c) reduce these networks to a low-dimensional set of mean-field equations.
Using the latter, we characterize the bifurcation structure of spiking neural networks across different parameter distributions, demonstrating both the critical role of parameter heterogeneity in neural network dynamics and how our method allows it to be studied systematically. 
As our method applies to any system with sufficiently many globally coupled phase oscillators that are governed by Eqs.~\eqref{eq:theta} and \eqref{eq:OA_coupling}, it provides a powerful approach for studying the impact of the oscillator-intrinsic parameter distribution $\rho(\omega)$on the the dynamics of such systems. 

\textit{Lorentzian Mixture Approach ---}
Extending an approach we previously established for spiking neural networks \cite{gast_mean-field_2021}, we approximate arbitrary distributions via a sum of Lorentzian distributions:
\begin{align}
    \rho(\omega) \approx \rho^*_M(\omega) = \frac{1}{\pi} \sum_{m=1}^M w_m \frac{\Delta_m}{(\omega-\bar \omega_m)^2 + \Delta_m^2}, \label{eq:lorentz_mix}
\end{align}
where $\bar \omega_m$ and $\Delta_m$ are the centers and half-widths of $M$ individual Lorentzian distributions, respectively, and $w_m$ are the scalar weights associated with each Lorentzian distribution that satisfy $w_m \geq 0.0$ and $\sum_m w_m = 1.0$.
For any set of i.i.d.\ samples $\omega_1, \omega_2, .., \omega_N$ from $\rho(\omega)$, we define the loss function as the Cramer-von Mises statistic $W^2$: 
\begin{equation}
    \mathcal{L}(F, F^*_M) = W^2 =  \int_{-\infty} ^{\infty} [F(\omega) - F^*_M(\omega)]^2 d F^*_M(\omega) \label{eq:loss},
\end{equation}
i.e. the squared difference between the cumulative distribution function (CDF) $F^*_M(\omega) = \int_{-\infty} ^{\omega} \rho^*_M(\omega) d \omega$ and the empirical CDF $F(\omega) = \frac{1}{N} \sum_{i=1}^N \mathbf{1}(\omega_i\leq\omega)$.
We treat the number of distributions $M$ as a hyperparameter, selected by greedy search: for each candidate $M$ we minimize Eq.~\eqref{eq:loss} via gradient descent; we then compute an outer loss $\mathcal{L}_{\mathrm{outer}}(M) = \mathcal{L}(F, F^*_M) + \lambda M$ for that $M$.
Starting from small $M$, we increment it until the outer loss plateaus or $M$ reaches a maximum allowed value $M_\textrm{max}$. 
The penalty $\lambda M$ in the outer loss discourages large $M$; thus the meta parameter $\lambda$ balances complexity of the Lorentzian mixture against fit accuracy.

Importantly, the Lorentzian mixture admits a closed-form CDF
\begin{align}
    F^*_M(\omega) &= \sum_{m=1}^M w_m \Omega_m(\omega),\\
    \Omega_m(\omega) &= \frac{1}{2} + \frac{1}{\pi} \arctan(\frac{\omega - \bar \omega_m}{\Delta_m}). \label{eq:cdf}
\end{align}
Gradient descent with respect to the parameters $\mathbf{u} = (w_1, w_2, ..., w_m, \bar \omega_1, \bar \omega_2, ..., \bar \omega_m, \Delta_1, \Delta_2, ..., \Delta_m)$ requires the evaluation of the partial derivatives
\begin{equation}
    \frac{\partial W^2}{\partial \mathbf{u}_i} = \frac{\partial W^2}{\partial F^*_M(\omega)} \frac{\partial F^*_M(\omega)}{\partial \mathbf{u}_i},
\end{equation}
which we also obtained analytically (see Appendix A).

Together, gradient descent minimization of Eq.~\eqref{eq:loss} can be performed with a fully analytic gradient, using Eq.~\eqref{eq:cdf} for the analytical Lorentzian mixture CDF.
Additional optimization constraints are $w_m \geq 0$ for all $m \in {1,2,..,M}$, $\sum_{m=1}^M w_m = 1$, and $\Delta_m \geq 0$ for all $m \in {1,2,..,M}$.
For a detailed description of the constrained optimization algorithm that we used throughout this letter, see Appendix B.

\textit{LMMF Equations for Globally Coupled Kuramoto Oscillators ---}
As a first demonstration of our method, we consider a network of globally coupled Kuramoto oscillators, i.e. Eq~\eqref{eq:theta} with $G(\theta_j, \theta_i) = \sin(\theta_j-\theta_i)$, where we select the distribution of intrinsic oscillator frequencies to be a mixture of two Gaussian distributions:
\begin{equation}
    \rho(\omega) = c_1 \mathcal{N}(\omega,\mu_1,\sigma_1) + c_2 \mathcal{N}(\omega,\mu_2,\sigma_2),
\end{equation}
with $c_1 + c_2 = 1.0$.

\begin{figure*}[ht!]
    \centering
    \includegraphics[width=1.0\linewidth]{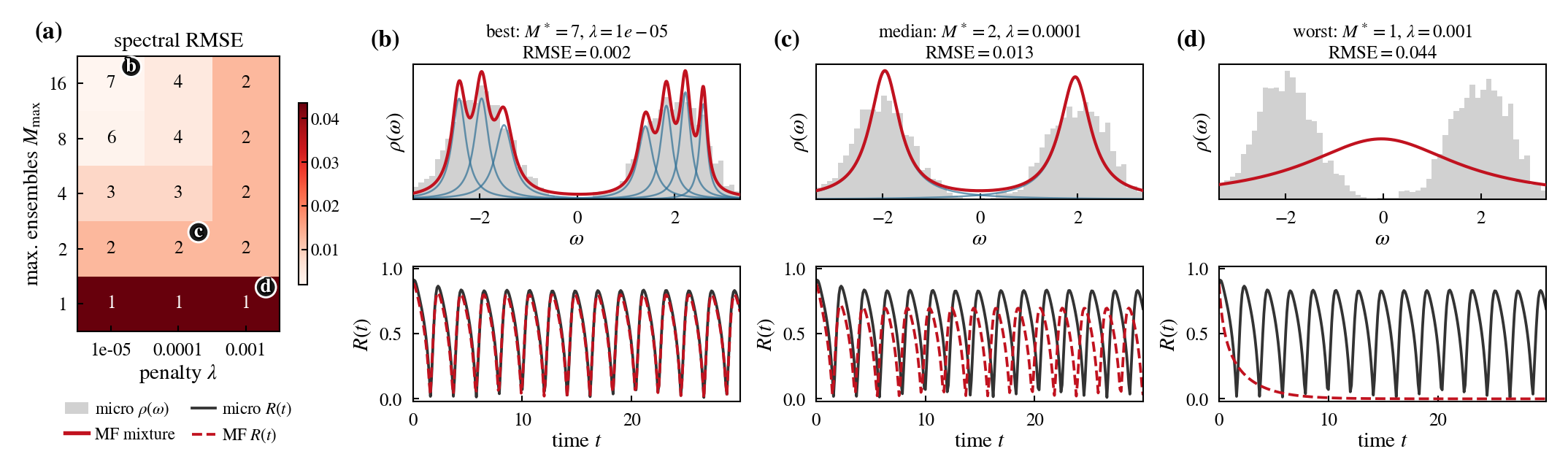}
    \caption{Lorentzian mixture mean-field (LMMF) model accurately captures the macroscopic dynamics of a network of Kuramoto oscillators with multimodal intrinsic frequency distribution. \textbf{(a)} LMMF model-fitting performance as a function of meta parameters $M_\textrm{max}$ and $\lambda$. The color represents the root mean square error between the Fourier transformed average phase coherence dynamics of the two models. The numbers in each cell represent the number of Lorentzian distributions $M$ that led to the best fit. \textbf{(b-d)} Example LMMF fits for different values of $M_{max}$ and $\lambda$. The top row depicts the fit to the empirical distribution of oscillator frequencies, whereas the bottom row depicts the average phase coherence dynamics of the Kuramoto oscillator network and the LMMF equations.}
    \label{fig:lorentz_mix}
\end{figure*}

Using our optimization algorithm, $\rho(\omega)$ can be approximated by a Lorentzian mixture $\rho^*_M(\omega)$.
If we consider each Lorentzian distribution of the mixture as a distinct ensemble of Kuramoto oscillators with a Lorentzian frequency distribution, then we may derive a system of mean-field equations using the Ott-Antonsen ansatz in the limit $N \rightarrow \infty$.
We obtain
\begin{align}
    \dot R_m &= -\Delta_m R_m + \frac{1-R_m^2}{2}K\sum_{l=1}^{M} w_l R_l \cos(\Psi_l - \Psi_m), \label{eq:r_m}\\
    \dot \Psi_m &= \Omega_m + \frac{1 + R_m^2}{2R_m} K \sum_{l=1}^M w_l R_l \sin(\Psi_l - \Psi_m), \label{eq:psi_m}
\end{align}
for the average phase coherence $R_m$ and the average phase $\Psi_m$ of each ensemble, where the mixture weights $w_m$ enter the equations as the relative contributions of each ensemble to the overall mean-field drive. 
We call this system the Lorentzian mixture mean-field (LMMF) model.

Comparing the dynamics of the mean-field Eqs.~\eqref{eq:r_m} and \eqref{eq:psi_m} to a system of $N=5000$ Kuramoto oscillators for different $M_{max}$ and penalties $\lambda$, we find that the LMMF equations can (i) faithfully capture the Kuramoto network dynamics, and (ii) achieve a considerable dimensionality reduction (see Fig.~\ref{fig:lorentz_mix}). 
Given sufficient flexibility in terms of $M_{max}$ and $\lambda$, the optimization procedure consistently stops at $M \approx 6-8$ across different random realizations of the empirical oscillator distribution.
Whereas $M=2$ suffices to predict the existence of the periodic solution that the Kuramoto system expresses (see Fig.~\ref{fig:lorentz_mix}c), $M \approx 6-8$ captures the average phase coherence dynamics of the finite-size Kuramoto system almost exactly.  
Note that the optimal $M$ depends on the choice of the meta parameter $\lambda$ (see Fig.~\ref{fig:lorentz_mix}a). 

\textit{Rational Frequency Distributions ---}
We next compare our approach to previous works that use an ensemble approach to capture the dynamics of globally coupled Kuramoto oscillator systems with a Gaussian frequency distribution \cite{campa_study_2022,skardal_low-dimensional_2018} (see also \cite{pyragas_meanfieldequations_2022,pyragas_mean-field_2024} for related work on quadratic integrate-and-fire neurons with Gaussian parameter distributions).
In \cite{campa_study_2022}, the authors uses a truncated power expansion to approximate the Gaussian distribution with a weighted sum of rational distributions.
They report a high accuracy in locating the synchronization threshold, but find a considerable difference in the transient dynamics at finite truncation orders.
In \cite{skardal_low-dimensional_2018}, Skardal extends the Ott-Antonsen ansatz to globally coupled Kuramoto oscillators with rational frequency distributions, showing that the reduced dynamics remain low-dimensional but require a number of equations that grows linearly with the distribution's order.

In our previous example, we found that our model captured the transient dynamics of the Kuramoto system with high accuracy.
To examine this property of our model more systematically, we therefore examined its performance on the rational frequency distributions studied by Skardal \cite{skardal_low-dimensional_2018}.
As in the previous section, we consider systems of globally coupled Kuramoto oscillators, but with a rational oscillator frequency distribution
\begin{equation}
    \rho(\omega) = g_n(\omega) = \frac{n\sin (\pi / 2n) \Delta^{2n-1}}{\pi(\omega^{2n} + \Delta^{2n})}, \label{eq:rational}
\end{equation}
which reduces to the Lorentzian distribution for $n=1$ and converges to the uniform distribution in the limit $n \rightarrow \infty$.

\begin{figure}[!ht]
    \centering
    \includegraphics[width=1.0\linewidth]{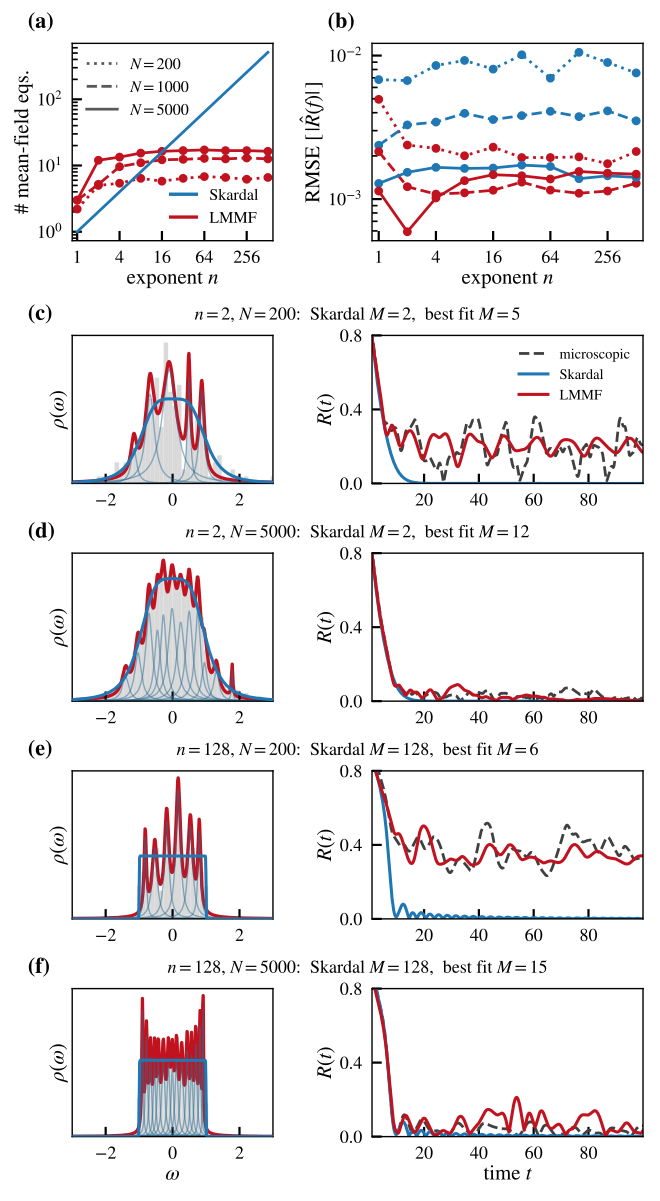}
    \caption{Low-dimensional LMMF equations capture finite size bias in Kuramoto oscillator networks with rational frequency distributions. \textbf{(a)} Number of complex equations in the LMMF model (red) and the Skardal mean-field model (blue) \cite{skardal_low-dimensional_2018}. \textbf{(b)} Performance of the LMMF equations vs. the Skardal mean-field equations in predicting the microscopic phase coherence dynamics. \textbf{(c-f)} Examples of oscillator frequency distribution fits (left column) and phase coherence dynamics (right column) for different sample sizes $N$ and distribution exponents $n$. In the left column, the blue line depicts the probability density $g_n(\omega)$ used by the Skardal model, the grey-filled bars depict the distribution of the microscopic samples drawn from $g_n(\omega)$, and the red line depicts the Lorentzian mixture fit of the sample distribution.}
    \label{fig:skardal}
\end{figure}

For different choices of $n$, we randomly sample individual oscillator frequencies $\omega_i$ from Eq.~\eqref{eq:rational}, and then compare the average phase coherence dynamics $R(t)$ of the Kuramoto system, the mean-field equations derived by Skardal \cite{skardal_low-dimensional_2018}, and our LMMF equations fitted to the distribution of samples $\omega_i$.
We consistently find that the LMMF equations capture $R(t)$ more closely than the Skardal mean-field equations (see Fig.~\ref{fig:skardal}).
Moreover, at high $n$ the LMMF equations are of considerably lower dimensionality than the Skardal equations, thus providing a more accurate representation of the Kuramoto system with a smaller number of mean-field equations (see Fig.~\ref{fig:skardal}b).

This result seems counterintuitive given the exact nature of the Skardal solution.
However, whereas the Skardal mean-field equations are fully accurate for the frequency density function given by Eq.~\eqref{eq:rational} as $N\rightarrow\infty$, even $N=5000$ randomly drawn samples $\omega_i$ from the density function \eqref{eq:rational} may introduce a finite size bias.
Our results suggest that the LMMF equations capture these small deviations of the empirical distribution in a way that better predicts the behavior of the finite-size system, despite the fact that the LMMF equations assume the limit $N_m \rightarrow \infty$.
Hence, we conclude that the LMMF equations are particularly well-suited to study the emergent dynamics of finite-size systems of globally coupled phase oscillators with heterogeneous oscillator frequencies.

\textit{Oscillator Heterogeneity ---}
Nodal heterogeneity, i.e. quenched disorder of parameters that control the behavior of nodes within large interacting networks, has been shown to play an important, non-trivial role for the dynamics of complex systems \cite{dahmen_how_2026}.
In coupled phase oscillators, for example, heterogeneity in oscillator frequencies can have either a synchronizing or a de-synchronizing effect, as well as promote or inhibit chimera states \cite{motter_network_2005,ashwin_extreme_2006,laing_chimera_2009,nishikawa_symmetric_2016,sugitani_synchronizing_2021}.
This has been shown to have direct implications for real-world problems that include interacting periodic processes, such as power grid stability \cite{molnar_asymmetry_2021}, active particle motion \cite{yang_emergent_2022}, and neural population coding \cite{padmanabhan_intrinsic_2010}.
As a final result, we report that our LMMF approach permits study of the role of parameter heterogeneity in globally coupled phase oscillator systems.

\begin{figure}[!ht]
    \centering
    \includegraphics[width=1.0\linewidth]{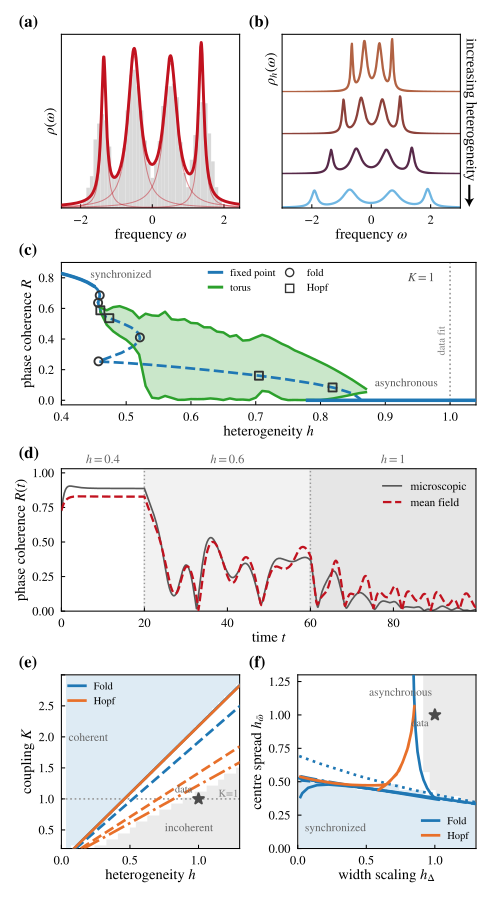}
    \caption{Global heterogeneity parameters for analyzing Lorentzian mixture fits. \textbf{(a)} The empirical distribution (grey) of $\omega_i$ and the Lorentzian mixture fit (red). \textbf{(b)} Effects of changes in $h = h_{\bar \omega} = h_{\Delta}$ on the Lorentzian mixture density. \textbf{(c)} 1D bifurcation diagram for the oscillator heterogeneity $h$. The steady-state solution branch (blue) and bifurcation points were obtained via parameter continuation of the synchronized fixed point solution of the LMMF equations, whereas the envelope of the quasi-periodic torus was approximated via numerical simulations of the LMMF equation dynamics in response to a slow, bidirectional ramp of $h$. \textbf{(d)} Average phase coherence dynamics of the LMMF equations and the Kuramoto oscillator network in response to steps in $h$. \textbf{(e-f)} 2D bifurcation diagrams of the Hopf and fold bifurcation loci in the $h-J$ and $h_{\bar \omega}-h_{\Delta}$ parameter spaces, respectively. Shaded regions represent the regions in parameter space where the Kuramoto oscillator network converged to an asynchronous or synchronized fixed point. }
    \label{fig:heterogeneity}
\end{figure}

We demonstrate our approach using the example of a mixture of four Gaussian distributions, and extend this to real neurophysiology data in Appendix C. 
We first sample $N=5000$ oscillator frequencies $\omega_i$ from the Gaussian mixture, and fit this empirical distribution with the Lorentzian mixture $\rho^*_M(\omega)$ (Fig.~\ref{fig:heterogeneity}a).
To study the effect of heterogeneity in this system, we then re-scale the centers and half-widths of the individual Lorentzian distributions, such that
\begin{align}
    \bar \omega_m^* &= h_{\bar \omega} \bar \omega_m + (1-h_{\bar \omega}) \sum_{l=1}^M w_l \bar \omega_l,\\
    \Delta_m^* &= h_{\Delta} \Delta_m.
\end{align}
This way, the LMMF approach allows for a systematic study of the system dynamics as a function of two global heterogeneity parameters: the distance $h_{\bar \omega}$ of the Lorentzian centers to the global distribution center, and the global scaling of the individual half-widths $h_{\Delta}$.
Choosing $h_{\bar \omega} = f_{\bar \omega}(h)$ and $h_{\Delta} = f_{\Delta}(h)$, the two can be studied as functions of a single, global heterogeneity parameter (see Fig.~\ref{fig:heterogeneity}b for the simple case $h_{\bar \omega} = h_{\Delta} = h$). 

Analyzing the fixed point solutions of the fitted LMMF equations with respect to changes in $h$ via the bifurcation analysis software PyCoBi \cite{gast_pyratescode-generation_2023}, we find that a stable, synchronized fixed point exists at small values of $h$, then loses stability through a fold bifurcation and eventually gives rise to a quasi-periodic torus via a supercritical Hopf bifurcation (see Fig.~\ref{fig:heterogeneity}c-d).
Tracing the loci of these codimension 1 bifurcations in the 2D parameter spaces spanned by $h-K$ and $h_{\Delta}-h_{\bar \omega}$, we obtain detailed mechanistic insight how distortions of the oscillator frequency distribution $\rho(\omega)$ affect the system dynamics (see Fig.~\ref{fig:heterogeneity}e-f).

\textit{Conclusion ---}
In this letter, we have introduced the Lorentzian mixture mean-field (LMMF) approach for studying the dynamics of coupled phase oscillator systems governed by arbitrary parameter distributions.
We have shown that our method can provide drastic dimensionality reductions in terms of the system equations, while still capturing the macroscopic dynamics of finite size oscillator systems accurately.
Finally, we have established a systematic method for studying the dynamic effects of complex distortions of the shape of parameter distributions within our LMMF framework.

\begin{acknowledgments}
\textit{Acknowledgments ---} We thank Juergen Kurths and Marcus Benna for helpful comments and inspirational discussions with respect to this manuscript.
This work was supported by NINDS Grant 1RF1NS132912, a Pew Biomedical Scholars award, and a McKnight Scholars award (awarded to AK).
HS acknowledges support by the Czech Science Foundation (project 25-15412L), the ERDF-Project Brain dynamics (CZ.02.01.01/00/22\_008/0004643), and a Lumina-Quaeruntur fellowship by the Czech Academy of Sciences (LQ100302301).
\end{acknowledgments}

\bibliography{references}

\appendix

\section*{End Matter}

\textit{Appendix A: Cramer-von Mises Loss Gradient \label{sec:app_gradient} ---}
In the following, we will provide the analytical derivation of the gradient with respect to all model parameters parameters $\mathbf{u}_i$. 
For ordered sets of $N$ i.i.d.\ samples $\omega_i$, the gradient with respect to the mixture weights $w_m$ is 
\begin{align}
    \frac{\partial W^2}{\partial w_m} &= \frac{\partial W^2}{\partial F^*_M(\omega)} \frac{\partial F^*_M(\omega)}{\partial w_m}\\
    &= 2 \sum_{i=1}^N \frac{\partial F^*_M(\omega_i)}{\partial w_m}[F(\omega_i)-F^*_M(\omega_i)]\\
    &= 2 \sum_{i=1}^N \Omega(\omega_i) [F(\omega_i)-F^*_M(\omega_i)],
\end{align}
with $\Omega(\omega_i)$ given by Eq.~\eqref{eq:cdf}.
The gradient with respect to the Lorentzian centers $\bar \omega_m$ is
\begin{align}
    \frac{\partial W^2}{\partial \bar \omega_m} &= \frac{\partial W^2}{\partial F^*_M(\omega)} \frac{\partial F^*_M(\omega)}{\partial \bar \omega_m}\\
    &= 2 \sum_{i=1}^N \frac{\partial F^*_M(\omega_i)}{\partial \bar \omega_m}[F(\omega_i)-F^*_M(\omega_i)]\\
    &= -2 w_m \sum_{i=1}^N \frac{\partial \Omega(\omega_i)}{\partial \bar \omega_m}[F(\omega_i)-F^*_M(\omega_i)],\\
    &= -2 w_m \sum_{i=1}^N \rho_m(\omega_i)[F(\omega_i)-F^*_M(\omega_i)],
\end{align}
with $\rho(\omega_i)$ given by Eq.~\eqref{eq:lorentz}.
For the gradient with respect to the Lorentzian half-widths, we obtain
\begin{align}
    \frac{\partial W^2}{\partial \Delta_m} &= \frac{\partial W^2}{\partial F^*_M(\omega)} \frac{\partial F^*_M(\omega)}{\partial \Delta_m}\\
    &= 2 \sum_{i=1}^N \frac{\partial F^*_M(\omega_i)}{\partial \Delta_m}[F(\omega_i)-F^*_M(\omega_i)]\\
    &= - 2 w_m \sum_{i=1}^N \frac{\partial \Omega(\omega_i)}{\partial \Delta_m}[F(\omega_i)-F^*_M(\omega_i)]\\
    &= - \frac{2 w_m}{\pi} \sum_{i=1}^N \frac{\omega_i-\bar \omega_m}{(\omega_i - \bar \omega_m)^2 + \Delta_m^2}[F(\omega_i)-F^*_M(\omega_i)].
\end{align}

\textit{Appendix B: Lorentzian Mixture Optimization Algorithm \label{sec:app_algorithm} ---}
Here, we present a nested 2-level optimization algorithm for the Lorentzian mixture given by Eq.~\eqref{eq:lorentz_mix}, where the outer level performs the selection of the optimal number of ensembles $M$ and the inner level minimizes Eq.~\eqref{eq:loss} for a given $M$ via gradient descent.  

The outer loop selects $M \in {1,...,M_{max}}$ via a greedy search with two early-stopping criteria, whereas the inner loop uses the Sequential Least Squares Programming (SLSQP) algorithm with multiple seeding points for gradient descent \cite{kraft1988software}, as it allows to implement boundary and equality constraints such that all solutions to the minimization problem $\min_{\mathbf{u}} \mathcal{L}(F, F^*_M(\mathbf{u}))$ obey $w_m \geq 0$ for all $m \in {1,2,..,M}$, $\sum_{m=1}^M w_m = 1$, and $\Delta_m \geq 0$ for all $m \in {1,2,..,M}$. 
Since the inner loop is minimizing the Cramer-von Mises loss $W^2$, we can make use of the test statistic $T = N W^2$ for the difference between the empirical distribution and the model distribution, which has been tabulated for finite $N$ \cite{anderson_distribution_1962}.
Starting from $M=1$, the outer optimization loop uses the SLSQP optimization to find the set of parameters $\mathbf{u}$ that minimizes Eq.~\eqref{eq:loss}.
The outer loop stops if $T < 1-\alpha$, i.e. if the empirical and model distributions are statistically indistinguishable at the significance level $\alpha$. 
As a second stopping criterion, the outer loss $\mathcal{L}_{\mathrm{outer}}(M) = \mathcal{L}_{\mathrm{inner}}(F, F^*_M) + \lambda M$ is used, where $\mathcal{L}_{\mathrm{inner}}(F, F^*_M)$ is the final loss after the SLSQP optimization for a given $M$.
If $\mathcal{L}_{\mathrm{outer}}(M+k) \leq \mathcal{L}_{\mathrm{outer}}(M)$ for $k=[1,...,K]$, the outer loop is stopped and the solution for the $M$ with the lowest $\mathcal{L}_{\mathrm{outer}}(M)$ is returned.

This concludes an algorithm for approximating arbitrary parameter distributions $\rho(\omega)$ via a Lorentzian mixture given by Eq.~\eqref{eq:lorentz_mix}.
It leverages the analytical properties of the Lorentzian distributions to fit the CDF of the Lorentizan mixture to empirical parameter distributions via the Cramer-von Mises loss given by Eq.~\eqref{eq:loss}, for which the gradients with respect to the parameters $\mathbf{u}$ can be obtained analytically. 
Finally, the algorithm exploits that the mixture combines all Lorentzian distributions linearly, allowing for a greedy search strategy with early stopping over the parameter $M$. 

\textit{Appendix C: Dynamics of Globally Coupled Spiking Neurons from Different Cortical Layers\label{sec:app_qif} ---}
The relationship between neural heterogeneity, synchronization of neural dynamics and population coding has been a particular subject of increasing research interest, not only within the field of neuroscience \cite{zeldenrust_efficient_2021,cembrowski_continuous_2018,mittal_resonating_2021,lafond-mercier_neural_2025,dahmen_how_2026}, but also in machine learning \cite{perez-nieves_neural_2021,winston_heterogeneity_2023} and analog computing \cite{payvand_self-organization_2022,zendrikov_brain-inspired_2023}.
A recent idea in this area has been that dynamic changes of neural heterogeneity -- caused by neuron-intrinsic adaptation processes or neuromodulatory signals -- could serve to control neurocomputational regimes \cite{papadopoulos_modulation_2025,dahmen_how_2026}.
While a range of studies examined the macroscopic dynamics of networks of spiking neurons as a function of the width of a parameter distribution \cite{mejias_optimal_2012,di_volo_optimal_2021,rich_loss_2022,hutt_intrinsic_2023,gast_neural_2024,papadopoulos_modulation_2025}, an application of these ideas to empirical data sets of neural heterogeneity is still missing.

Here, we present the LMMF approach as a novel tool for (i) establishing a direct connection between neural network models and empirical data on neural heterogeneity, and (ii) the systematic study of changes in empirical parameter distributions. 
Using open-source electrophysiological provided by the Allen Brain Institute for cortical neurons in mice \cite{gouwens2019,allen_ephys_whitepaper,allen_celltypes_db}, we obtained empirical distributions of the distance-to-threshold $D_i = V_{\theta,i} - V_{r,i}$, where $V_{\theta,i}$ and $V_{r,i}$ refer to the spike threshold and resting membrane potential of the $i^{\mathrm{th}}$ neuron in a data set, respectively.
Specifically, we extracted empirical distributions of $D_i$ for fast-spiking interneurons in cortical layers 2/3 ($N=46$) and 5/6 ($N=126$) available through the Allen Software Development Kit \cite{allensdk}.
Using the LMMF approach, we would like to study how differences in the distributions between cortical layers translate to differences in the population dynamics. 
To this end, we consider QIF networks of the form
\begin{align}
    \dot V_i &= (V_i - V_r)(V_i - V_{\theta,i}) + J s + I(t), \label{eq:v_i}\\
    s &= \frac{1}{N}\sum_{i=1}^N \int_0^{\infty} g(\tau) \sum_{k \backslash t_i^k<t-\tau} \delta(t-\tau-t_i^k) \text{d} \tau, \label{eq:s}
\end{align}
where $V_i$ is the membrane potential of neuron $i$, $t_i^k$ is the time of the $k^{th}$ spike emitted by neuron $i$, and $s$ is a convolution of the global spike rate with the synaptic response kernel $g(s)$.
A neuron emits a spike when $V_i \geq V_p$, i.e. when its membrane crosses a peak value, after which the membrane potential is reset to $V_i \leftarrow V_0$. 
To inform the QIF model Eqs.~\eqref{eq:v_i} and \eqref{eq:s} by the empirical distributions, we choose a global resting membrane potential $V_r = \bar V_r$ and defined $V_{\theta,i} = \bar V_r + D_i$, thus collapsing heterogeneity in $V_{r,i}$ and $V_{\theta,i}$ into a single parameter.

\begin{figure}[!ht]
    \centering
    \includegraphics[width=1.0\linewidth]{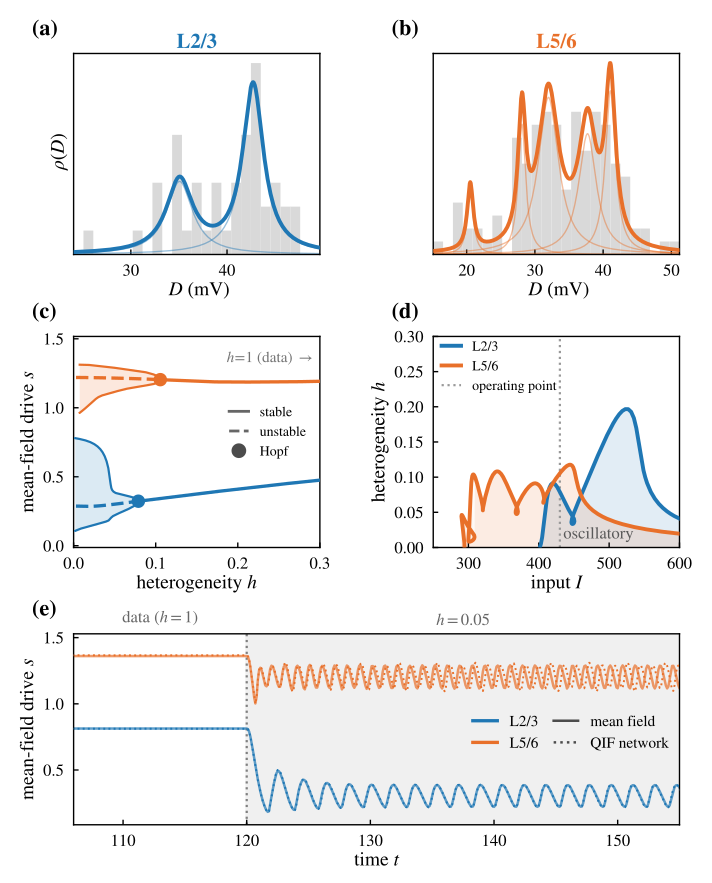}
    \caption{LMMF equations reveal heterogeneity-related synchronization of fast-spiking interneurons. \textbf{(a-b)} Lorentzian mixture fits to the empirical distributions of $D_i$ in layer 2/3 and layer 5/6 fast-spiking interneurons, respectively. \textbf{(c)} 1D bifurcation diagram of the globally coupled interneuron populations for $J=-100$ and $I=430$. \textbf{(d)} 2D bifurcation diagram depicting the Hopf bifurcation curves in the $I-h$ parameter space. \textbf{(e)} Synaptic activation dynamics of the mean-field equations and corresponding QIF networks in response to a decrease in neural heterogeneity.}
    \label{fig:pv_interneurons}
\end{figure}

We then used our LMMF approach to obtain a low-dimensional set of mean-field equations for the QIF network.
First, we fitted the Lorentzian mixture to the empirical distributions of $D_i$.
Then, we followed the derivation described in \cite{gast_macroscopic_2023} to obtain a set of mean-field equations for each QIF ensemble $m$ of the Lorentzian mixture:
\begin{align}
    \dot r_m &= \frac{\Delta_m \sigma_m}{\pi}(v_m-\bar V_r) + r_m(2 v_m - \bar V_r - \bar V_{\theta,m}), \label{eq:rate_m}\\
    \dot v_m &= (v_m-\bar V_r)(v_m-\bar V_{\theta,m}) + J s + I(t) \\ &- \pi r_m(\Delta_m \ \sigma_m + \pi r_m), \label{eq:v_m}\\
    s &= \sum_{m=1}^M w_m \int_0^{\infty} g(\tau) r_m(t-\tau) \text{d} \tau, \label{eq:s_m}
\end{align}
where $\sigma_m = \text{sign}(v_m - \bar V_r)$ reflects a state-dependent pole switch for solving the integral over the Lorentzian distribution
\begin{equation}
    \rho(V_{\theta,m}) = \frac{1}{\pi} \frac{\Delta_m}{[V_{\theta,m} - \bar V_{\theta,m}]^2 + \Delta_m^2}.
\end{equation}
The mean-field derivation builds on previous results, showing that globally coupled QIF neurons permit the application of the Ott-Antonsen ansatz in the limits $V_p \rightarrow \infty$ and $V_0 \rightarrow -\infty$ \cite{luke_complete_2013,montbrio2015macroscopic,pietras_exactfiringrate_2019,bick_understanding_2020,gast_mean-field_2020,gast_mean-field_2021,pietras_low-dimensional_2025}.
Using the LMMF equations, we analyzed the equilibria of the QIF system, their stability, as well as their bifurcation structure via the parameter continuation software \textit{PyCoBi} \cite{gast_pyratescode-generation_2023}.

We find subtle differences in the neural synchronization properties of fast-spiking interneurons in layer 2/3 vs. layer 5/6, which can be explained by the differences in their average distance-to-threshold (see Fig.~\ref{fig:pv_interneurons}a-b). 
Generally, both distributions reflect a high degree of neural heterogeneity that favor asynchronous neural dynamics; only at a substantial collapse in heterogeneity (parameterized through $h$) does the asynchronous fixed point loose stability via a supercritical Hopf bifurcation, giving rise to limit cycle oscillations (see Fig.~\ref{fig:pv_interneurons}c-e).

\begin{figure}
    \centering
    \includegraphics[width=1.0\linewidth]{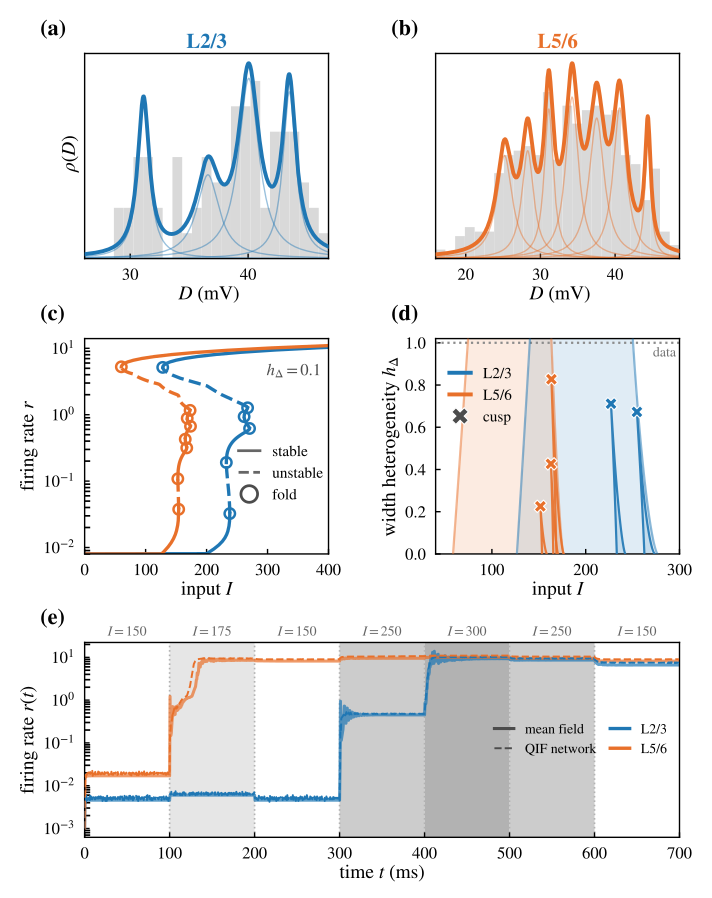}
    \caption{LMMF equations reveal heterogeneity-related multi-stability of pyramidal cells. \textbf{(a-b)} Lorentzian mixture fits to the empirical distributions of $D_i$ in layer 2/3 and layer 5/6 pyramidal cells, respectively. \textbf{(c)} 1D bifurcation diagram of the pyramidal cell populations for $J=100$. \textbf{(d)} 2D bifurcation diagram depicting the fold curves in the $I-h_{\Delta}$ parameter space. \textbf{(e)} Average firing rate dynamics of the mean-field equations and corresponding QIF networks in response to a stepping input.}
    \label{fig:pyramical_cells}
\end{figure}

Following the same procedure as described for the fast-spiking interneurons, we obtained $D_i$ for excitatory pyramidal cells of cortical layers 2/3 ($N=75$) vs. 5/6 ($N=488$) via the Allen Software Development Kit \cite{allensdk}.
Again, we analyzed the equilibria of recurrently coupled QIF neurons of the pyramidal cell type governed by those distributions by fitting the LMMF model to the empirical distributions and identifying the equilibria and bifurcations of Eqs.~\eqref{eq:rate_m}-\eqref{eq:s_m}.
We find differences in the input-output steady-state curves of layer 2/3 vs. layer 5/6 pyramidal cells, including the loci of fold bifurcations in the $J-I$ parameter space (see Fig.~\ref{fig:pyramical_cells}).
The more uniform distribution of $D_i$ in the layer 5/6 pyramidal cells requires less external input for the population to transition into an activated state in contrast to the multi-modal distribution of the layer 2/3 pyramidal cells. 
At low values of $h_{\Delta}$, multi-stable dynamic regimes exist due to the multi-modal distributions over $D_i$, allowing extrinsic inputs to switch between multiple stable firing rate configurations (see Fig.~\ref{fig:pyramical_cells}c-d).
Increased heterogeneity around the centers $\bar V_{\theta,m}$, implemented through $h_{\Delta}$, reduce the number of co-existing stable states.

These findings confirm previous results on the role of neural heterogeneity for multi-stability and neural synchronization \cite{gast_macroscopic_2023,hutt_intrinsic_2023,gast_neural_2024,papadopoulos_modulation_2025,dahmen_how_2026} and shows that the LMMF approach can reveal such heterogeneity effects in empirically measured parameter distributions with complex shapes (see Fig.~\ref{fig:pyramical_cells}a-b).

\end{document}